\documentclass[pra,english,reprint,nofootinbib,aps,superscriptaddress,showpacs,showkeys]{revtex4-2}

\usepackage{babel,calc,amsmath,amsthm,amssymb,graphicx,subfigure,xcolor,comment,times}
\usepackage{mathdots}
\usepackage[T1]{fontenc}
\usepackage[unicode=true]{hyperref}
\usepackage{booktabs}
\usepackage{mathtools}
\usepackage{braket}
\usepackage{dsfont}

\definecolor{linkblue}{rgb}{0, 0, 1}

\usepackage[normalem]{ulem}
\hypersetup{
	colorlinks=true,       		
	linkcolor=linkblue,          	
	citecolor=linkblue,             
	urlcolor=linkblue,          
}

\newtheorem*{theorem*}{Theorem}

\newtheorem*{corollary*}{Corollary}

\newtheorem*{lemma*}{Lemma}

\newtheorem*{proposition*}{Proposition}
\theoremstyle{definition}

\newtheorem*{definition*}{Definition}
\theoremstyle{remark}

\newtheorem*{remark*}{Remark}

\begin{document}
	\renewcommand{\figurename}{Fig.}
	
	\title{Reconstructing the multiphoton spatial wave function with coincidence wavefront sensing}
	
	\author{Yi~Zheng}
	\affiliation{CAS Key Laboratory of Quantum Information, University of Science and Technology of China, Hefei 230026, People's Republic of China}
	\affiliation{CAS Center For Excellence in Quantum Information and Quantum Physics, University of Science and Technology of China, Hefei 230026, People's Republic of China}
	
	\author{Mu~Yang}
	\affiliation{CAS Key Laboratory of Quantum Information, University of Science and Technology of China, Hefei 230026, People's Republic of China}
	\affiliation{CAS Center For Excellence in Quantum Information and Quantum Physics, University of Science and Technology of China, Hefei 230026, People's Republic of China}
	
	\author{Yu-Wei~Liao}
	\affiliation{CAS Key Laboratory of Quantum Information, University of Science and Technology of China, Hefei 230026, People's Republic of China}
	\affiliation{CAS Center For Excellence in Quantum Information and Quantum Physics, University of Science and Technology of China, Hefei 230026, People's Republic of China}
	
	\author{Jin-Shi~Xu}
	\email{jsxu@ustc.edu.cn}
	\affiliation{CAS Key Laboratory of Quantum Information, University of Science and Technology of China, Hefei 230026, People's Republic of China}
	\affiliation{CAS Center For Excellence in Quantum Information and Quantum Physics, University of Science and Technology of China, Hefei 230026, People's Republic of China}
	\affiliation{Hefei National Laboratory, University of Science and Technology of China, Hefei 230088, People's Republic of China}
	
	\author{Chuan-Feng~Li}
	\email{cfli@ustc.edu.cn}
	\affiliation{CAS Key Laboratory of Quantum Information, University of Science and Technology of China, Hefei 230026, People's Republic of China}
	\affiliation{CAS Center For Excellence in Quantum Information and Quantum Physics, University of Science and Technology of China, Hefei 230026, People's Republic of China}
	\affiliation{Hefei National Laboratory, University of Science and Technology of China, Hefei 230088, People's Republic of China}
	
	\author{Guang-Can~Guo}
	\affiliation{CAS Key Laboratory of Quantum Information, University of Science and Technology of China, Hefei 230026, People's Republic of China}
	\affiliation{CAS Center For Excellence in Quantum Information and Quantum Physics, University of Science and Technology of China, Hefei 230026, People's Republic of China}
	\affiliation{Hefei National Laboratory, University of Science and Technology of China, Hefei 230088, People's Republic of China}
	
	\date{\today}
	
	\begin{abstract}
		The quantum wave function of multiple particles provides additional information which is inaccessible to detectors working alone. Here, we introduce the coincidence wavefront sensing (CWS) method to reconstruct the phase of the multiphoton transverse spatial wave function. The spatially resolved coincidence photon counting is involved. Numerical simulations of two-photon cases using the weak measurement wavefront sensor are performed to test its correctness, and the phase information hidden in the correlation is revealed. Our work provides a direct spatial way to characterize multipartite quantum systems, and leads to fundamental studies like experimental Bohmian mechanics and applications in quantum optical technologies.
	\end{abstract}
	
	\maketitle
	
	\section{Introduction}
	
	As an essential part of quantum physics, the wave function is a complex quantity whose phase information is relatively difficult to measure. For photons, it may describe the transverse spatial mode of light under certain conditions \cite{photonwf,Lundeen11}. Direct and indirect measurement can be applied to detect the single-photon wave function. Direct measurement means it is obtained without special algorithms, and often involves the concept of weak value \cite{AAV,wvrmp} which is retrieved by either weak \cite{Lundeen11,Shi2015} or strong \cite{Zhang2020} measurements. Indirect measurement methods are mainly wavefront sensing techniques, including the celebrated Shack--Hartmann wavefront sensor (SHWS) \cite{SHWS} and a weak measurement wavefront sensor (WMWS) developed by our group \cite{Yang2020,Zheng2021,Zheng2022}, whose idea originates from an experimental setup \cite{Kocsis2011} to measure the transverse Bohmian velocity \cite{Bohm1,Bohm2} of photons. They are indirect because the directly measured quantity is the phase gradient of the wave function, and line integral algorithms are required to reconstruct the phase distribution \cite{SHWSrecon}.
	
	The joint wave function of $n$ particles $\psi(x_1,x_2,\ldots,x_n)$ provides additional correlation information which is undetectable when measuring only part of them. Photons entangled in the polarization degree of freedom (DOF) are relatively easier to measure jointly, such as using quantum state tomography with coincidence counting \cite{tomo,tomo2} or some direct measurement methods \cite{Pan2019,Chen2021}. There are also reports of measurement of the biphoton temporal or spectral wave function \cite{temp1,temp2}. Here, we consider the spatial DOF, which was discussed by Einstein, Podolsky and Rosen (EPR) in 1935 \cite{EPR}. Photon pairs generated by spontaneous parametric down-conversion (SPDC) \cite{corrreview} approximately have this form of state, where the positions of two particles are correlated and their momenta are anticorrelated. The joint probability distribution measurement in the position and momentum bases has been demonstrated \cite{Howell2004,Black2019}, and this correlation has been applied to ghost imaging \cite{ghostimag}. As for the phase or whole wave function measurement, previous works include modal decomposition \cite{Law2004,OAMtomo}, diffraction \cite{diffract}, interference with a reference beam \cite{refint}, and phase-shifting holography using entangled polarization DOF \cite{polarent}. Modal decomposition requires a basis like the Laguerre--Gaussian modes. In order to obtain a detailed result, a high truncation number is required and the state tomography is extremely complicated. Other methods are all limited to a given form of correlation. Hence, methods to retrieve the joint phase distribution from the position space can be more promising.
	
	In this work, we introduce the coincidence wavefront sensing (CWS) method, which extends wavefront sensing to multiple photons by putting a wavefront sensor on each path, and show that the two-photon joint spatial wave function can be reconstructed with cameras able to perform spatially resolved coincidence counting. Then, for numerical simulation, we generate photon count data using WMWS according to two types of wave functions, and design an algorithm to reconstruct the phase distribution, as the traditional zonal and modal method \cite{SHWSrecon} require large matrix operations, which is quite difficult for the four-dimensional (4D) function $\phi(x_1,y_1,x_2,y_2)$. Finally, we discuss the applications of the CWS method and the coincidence counting technique required for an experimental realization.
	
	\section{Theory}
	
\begin{figure}[t]
		\centering
		\includegraphics[width=.48\textwidth]{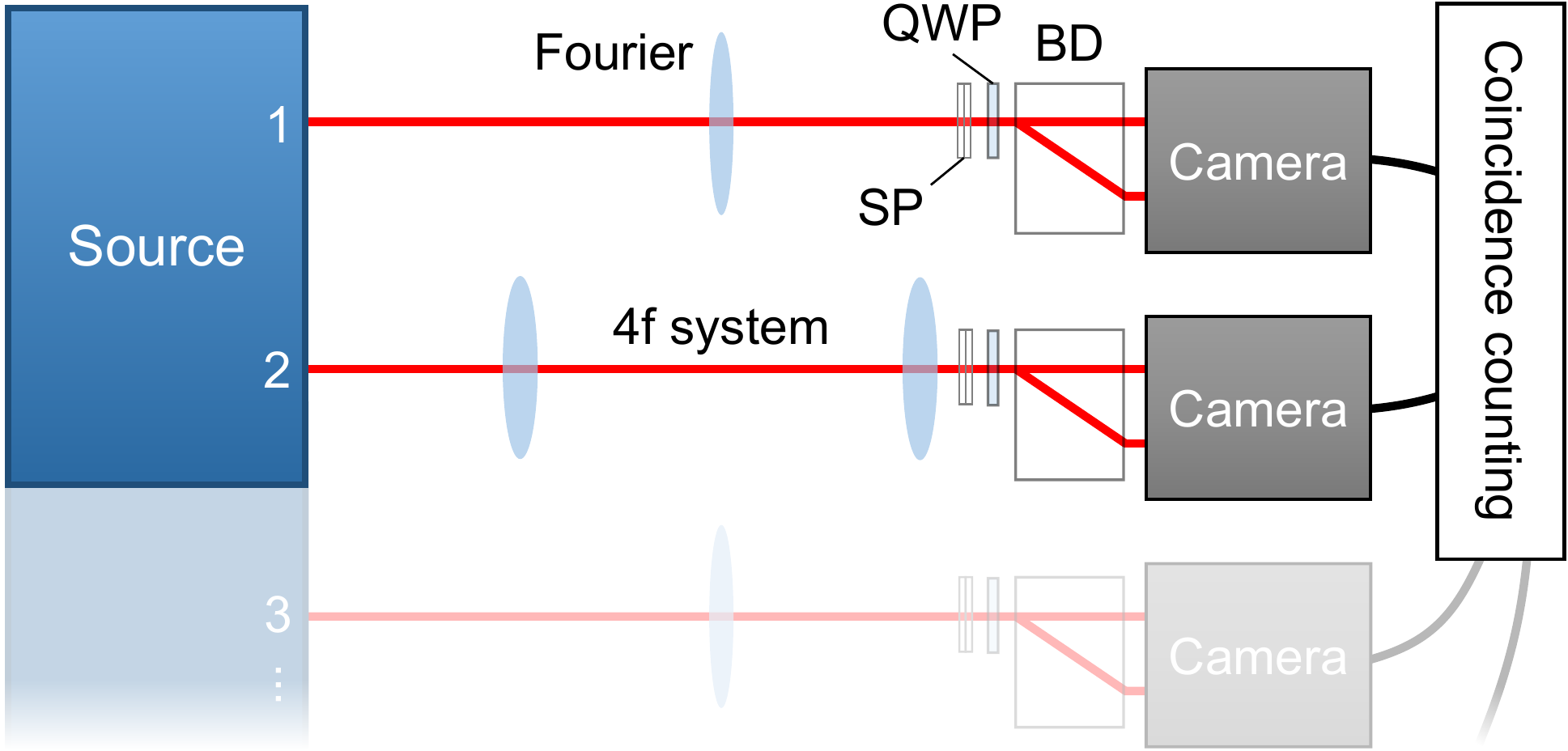}
		\caption{The scheme of the two-photon coincidence wavefront sensing with the weak measurement wavefront sensor. The source emits photon pairs with the state $|\psi\rangle|H_1\rangle|H_2\rangle$. A Fourier lens is used on path 1 to project the light field into the momentum space, and a $4f$ system is used on path 2 to image the field to the camera sensors. The weak measurement wavefront sensor consists of a Savart plate (SP), a quarter-wave plate (QWP), a beam displacer (BD) and a camera capable of two-photon coincidence counting. This setup is in principle extensible to $n$ photons.
		}
		\label{figsetup}
	\end{figure}

	In our proposed experimental setup as shown in Fig.~\ref{figsetup}, a light source emits photon pairs at the pure state $|\psi\rangle|H_1\rangle|H_2\rangle$, where $H$ denotes horizontal polarization, and $|\psi\rangle$ is a simplified form for two-photon spatial state \cite{quanoptnote}. On each path is a WMWS which consists of a Savart plate (SP, a combination of two identical thin birefringent crystals which are rotated by $90^\circ$ with respect to each other), a quarter-wave plate (QWP), a beam displacer (BD, a thick birefringent crystal, e.g., calcite) and a camera \cite{Zheng2022}. The distance between the camera sensor and the source is $d$. A pair of Fourier lenses (whose focal length $f=d/4$, forming a $4f$ system) or one Fourier lens ($f=d/2$; photons on the path are projected into the momentum space) is placed on each path. See Discussion for the choice of the two setups. In Fig.~\ref{figsetup}, one Fourier lens is on path 1, and the $4f$ system is on path 2. Considering the diffraction, the transverse wave function we measure is the photons propagated to the location of the camera sensor when the SP, the QWP, and the BD are absent \cite{Kocsis2011,Zheng2021,Zhu2021}. We denote the spatial state after Fourier transformation (FT) and inversion as $|\tilde{\psi}\rangle$, where
	\begin{equation}\label{fourier1}
		\tilde{\psi}(\mathbf{r}_1,\mathbf{r}_2)=\int d\mathbf{r}'_1 \psi(\mathbf{r}'_1,-\mathbf{r}_2)\exp\left(-i\frac{2\pi}{\lambda f}\mathbf{r}_1\cdot\mathbf{r}'_1\right),
	\end{equation}
	$f=d/2$, and $\lambda$ is the wavelength of light. The original wave function can be calculated by inverse FT once we obtain $\tilde{\psi}(\mathbf{r}_1,\mathbf{r}_2)$.
	
	This method needs two measurement steps, which we refer to as the $k_x$ and $k_y$ measurement. Letting $\tilde{\psi}(\mathbf{r}_1,\mathbf{r}_2)=A(\mathbf{r}_1,\mathbf{r}_2)\exp[i\phi(\mathbf{r}_1,\mathbf{r}_2)]$, and the phase gradient of the $j$th ($j=1,2$) photon $\mathbf{k}_j(\mathbf{r}_1,\mathbf{r}_2)=\nabla_j\phi=(k_{jx},k_{jy})$, in the $k_x$ measurement, the first crystal of the SP displaces diagonally ($D$) polarized light by $\mathbf{l}_+=(\mathbf{e}_x+\mathbf{e}_y)l$, and the second crystal displaces antidiagonally ($A$) polarized one by $\mathbf{l}_-=(-\mathbf{e}_x+\mathbf{e}_y)l$, where $\mathbf{e}_x,\mathbf{e}_y$ are unit vectors and $l$ is a small quantity for weak measurement. Then the $k_{1x},k_{2x}$ distribution will be obtained. In the $k_y$ measurement, both SPs are rotated counterclockwise by $90^\circ$, and $k_{1y},k_{2y}$ are obtained. Defining $|D\rangle=\frac{1}{\sqrt{2}}(|H\rangle+|V\rangle),|A\rangle=\frac{1}{\sqrt{2}}(|H\rangle-|V\rangle)$, and the left- ($L$) and right- ($R$) handed circular polarization $|L\rangle=\frac{1}{\sqrt{2}}(|H\rangle+i|V\rangle),|R\rangle=\frac{1}{\sqrt{2}}(|H\rangle-i|V\rangle)$, where $V$ denotes the vertical polarization, we use the $k_x$ measurement as an example. The overall quantum state after the SPs becomes (unnormalized)
	\begin{align}
		&|\tilde{\psi}_{++}\rangle|D_1\rangle|D_2\rangle+|\tilde{\psi}_{+-}\rangle|D_1\rangle|A_2\rangle\nonumber\\
		&+|\tilde{\psi}_{-+}\rangle|A_1\rangle|D_2\rangle+|\tilde{\psi}_{--}\rangle|A_1\rangle|A_2\rangle,
	\end{align}
	where $\langle\mathbf{r}_1,\mathbf{r}_2|\tilde{\psi}_{\pm_1\pm_2}\rangle=\tilde{\psi}(\mathbf{r}_1-\mathbf{l}_{\pm_1},\mathbf{r}_2-\mathbf{l}_{\pm_2})$ are the displaced wave functions. (The ``$\pm_i$'', ``$\mp_i$'', or ``${^L_R}_i$'' symbols with the same subscript are all replaced by either its upper or lower symbol.) Then the QWP and the BD separate the light fields with $L$ and $R$ polarization. From coincidence counting, we obtain the four joint intensity (or probability) distributions of the two photons
	\begin{equation}\label{intdist}
		I_{{^L_R}_1{^L_R}_2}(\mathbf{r}_1,\mathbf{r}_2)\propto\big|\tilde{\psi}_{\pm_1\pm_2}-\tilde{\psi}_{\mp_1\mp_2}+i\tilde{\psi}_{\pm_1\mp_2}+i\tilde{\psi}_{\mp_1\pm_2}\big|^2.
	\end{equation}
	The conditional intensity distributions are calculated by
	\begin{gather}
		I_{1{^L_R}}(\mathbf{r}_1|\mathbf{r}_2)=I_{{^L_R}L}(\mathbf{r}_1,\mathbf{r}_2)+I_{{^L_R}R}(\mathbf{r}_1,\mathbf{r}_2),\nonumber\\
		I_{2{^L_R}}(\mathbf{r}_2|\mathbf{r}_1)=I_{L{^L_R}}(\mathbf{r}_1,\mathbf{r}_2)+I_{R{^L_R}}(\mathbf{r}_1,\mathbf{r}_2).
	\end{gather}
	Defining the similar $A_{\pm_1\pm_2},\phi_{\pm_1\pm_2}$ notation, and taking the first-order approximations (see Appendix \ref{appA} for more details), we have
	\begin{align}\label{condintres}
		I_{j{^L_R}}(\mathbf{r}_j|\mathbf{r}_\textrm{other})\propto~ &A^2(\mathbf{r}_1-l\mathbf{e}_y,\mathbf{r}_2-l\mathbf{e}_y)\nonumber\\
		&\times\left\{1\mp\sin[2lk_{jx}(\mathbf{r}_1-l\mathbf{e}_y,\mathbf{r}_2-l\mathbf{e}_y)]\right\},
	\end{align}
	and thus
	\begin{equation}
		k_{jx}(\mathbf{r}_1-l\mathbf{e}_y,\mathbf{r}_2-l\mathbf{e}_y)\approx\frac{1}{2l}\arcsin\frac{I_{jR}-I_{jL}}{I_{jR}+I_{jL}}.
	\end{equation}
	In the $k_y$ measurement, $\mathbf{l}_+$ is redefined as $-(\mathbf{e}_x+\mathbf{e}_y)l$, and the formula is
	\begin{equation}
		k_{jy}(\mathbf{r}_1+l\mathbf{e}_x,\mathbf{r}_2+l\mathbf{e}_x)\approx\frac{1}{2l}\arcsin\frac{I_{jL}-I_{jR}}{I_{jL}+I_{jR}}.
	\end{equation}
	The phase gradient is related to the real part of the weak value of photonic transverse momentum
	\begin{equation}\label{weakvalueeqn}
		\mathbf{k}_j(\mathbf{r}_1,\mathbf{r}_2)=\frac{1}{\hbar}\operatorname{Re}\langle\hat{\mathbf{p}}_j\rangle_\textrm{w}=\frac{1}{\hbar}\operatorname{Re}\frac{\langle\mathbf{r}_1,\mathbf{r}_2|\hat{\mathbf{p}}_j|\tilde{\psi}\rangle}{\langle\mathbf{r}_1,\mathbf{r}_2|\tilde{\psi}\rangle},
	\end{equation} 
	where $\hbar$ is the reduced Planck constant. See Appendix \ref{appB} for a derivation using the framework of weak measurement. As we have all the partial derivatives of $\phi(\mathbf{r}_1,\mathbf{r}_2)$, it can now be reconstructed from the line integral \cite{SHWSrecon}
	\begin{equation}\label{lineint}
		\phi(\mathbf{r}_1,\mathbf{r}_2)=\int_{(\mathbf{0},\mathbf{0})}^{(\mathbf{r}_1,\mathbf{r}_2)}\mathbf{k}_1(\mathbf{r}'_1,\mathbf{r}'_2)\cdot d\mathbf{r}'_1+\mathbf{k}_2(\mathbf{r}'_1,\mathbf{r}'_2)\cdot d\mathbf{r}'_2,
	\end{equation}
	and the amplitude can be approximated using
	\begin{equation}
		A(\mathbf{r}_1-l\mathbf{e}_y,\mathbf{r}_2-l\mathbf{e}_y)\approx \sqrt{I_{1L}(\mathbf{r}_1|\mathbf{r}_2)+I_{1R}(\mathbf{r}_1|\mathbf{r}_2)}
	\end{equation}
	in the $k_x$ measurement. This scheme is, in principle, extensible to $n$ photons \cite{nphoton} (see Appendix \ref{appC} for more details).
	
	If the WMWS detects each path individually, the phase derivative we measure becomes the weak value when the initial state is mixed, described by the reduced density operator \cite{wvmix}, corresponding to the mutual coherence function in classical optics \cite{Zheng2021,Zhou2021,qrmcf}. The phase correlation information is generally lost. Additionally, if the two photons have good spatial correlation, the individual states are almost spatially incoherent. When a phase is added on one path (for example, $|\tilde{\psi}'\rangle=\exp[i\phi_\textrm{add}(\hat{\mathbf{r}}_2)]|\tilde{\psi}\rangle$), it is nearly undetectable only from path 2. See Appendix \ref{appD} for more details.
	
	\section{Numerical Simulation}\label{sec3}
	
	We developed a computer program for numerical simulation \cite{SourceCode}. Given the wave function $\tilde{\psi}(\mathbf{r}_1,\mathbf{r}_2)$, the theoretical joint probability distributions from Eq.~\eqref{intdist} are used for the Monte Carlo method to generate a number of point pairs. Then, $k_{jx}(\mathbf{r}_1,\mathbf{r}_2)$ and $k_{jy}(\mathbf{r}_1,\mathbf{r}_2)$ are calculated from the generated joint intensity distributions $I_{{^L_R}_1{^L_R}_2}(\mathbf{r}_1,\mathbf{r}_2)$.
	
	We designed an algorithm (see Appendix \ref{appE} for details) to reconstruct $\phi(\mathbf{r}_1,\mathbf{r}_2)$ in the continuous region within the region of interest (ROI) where the total joint intensity distributions $I_{1L}+I_{1R}$ in the $k_x$ and $k_y$ measurement are not zero. Then, we set $\lambda=800~\textrm{nm}$, $f=20~\textrm{cm}$, $l=25~\textrm{\textmu m}$, the camera pixel width $\varDelta=25~\textrm{\textmu m}$, and the ROI to be a square of $2.5~\textrm{mm}\times2.5~\textrm{mm}$ ($100\times100$ pixels), and perform the simulation for two cases.
	
	\begin{figure}[b]
		\centering
		\includegraphics[width=.48\textwidth]{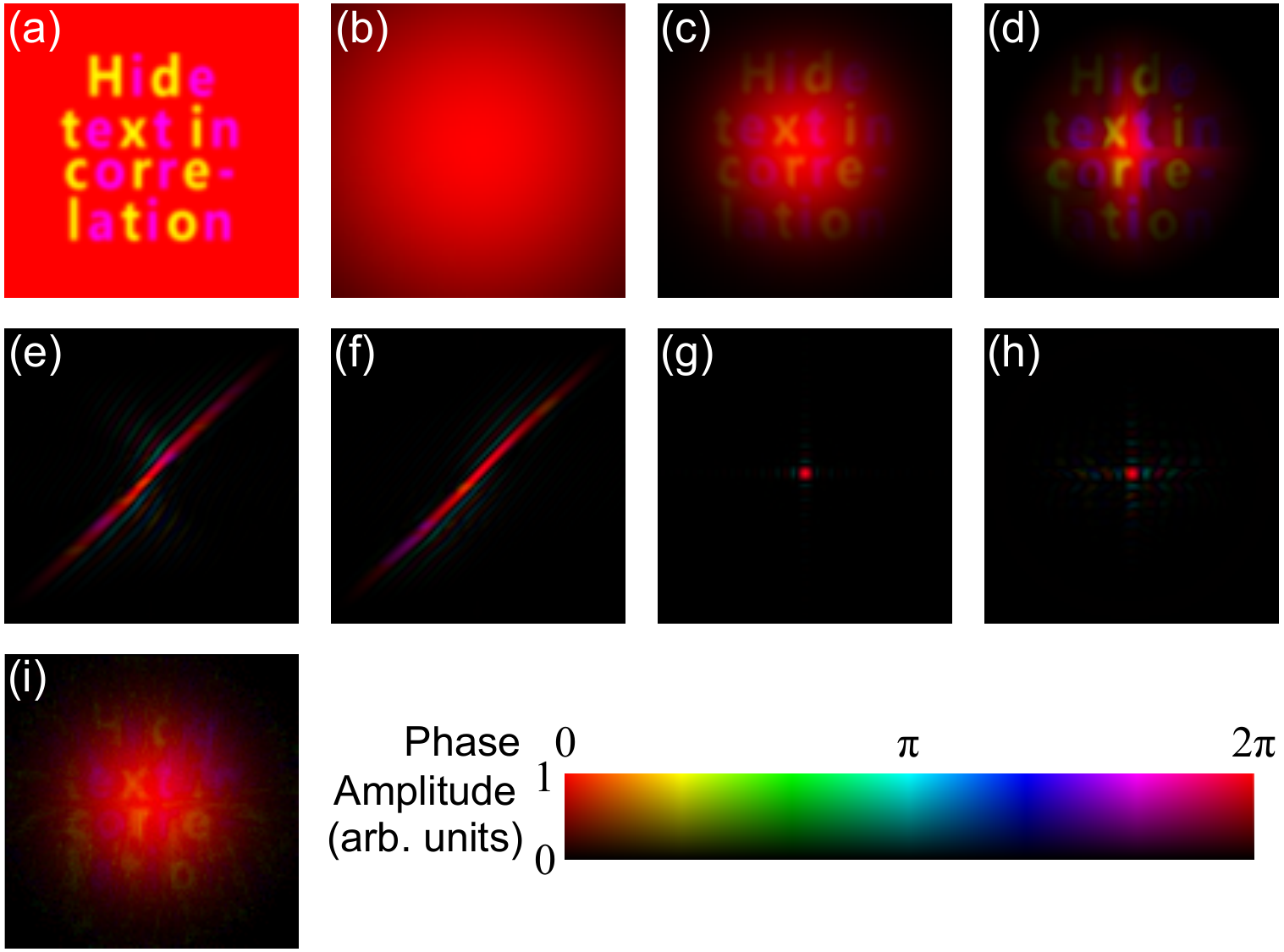}
		\caption{The two-photon wave function reconstruction result of the first case of CWS numerical simulation (see text), where photons have a good correlation and a phase is added on path 2. (a) The phase pattern. (b) and (c) The marginal reconstruction of path 1 and 2 respectively. Slices of the reconstructed wave function: (d) $\psi_\textrm{rec}(x,y,x,y)$. (e) $\psi_\textrm{rec}(x,0,y,0)$. (f) $\psi_\textrm{rec}(0,x,0,y)$. (g) $\psi_\textrm{rec}(x,y,0,0)$. (h) $\psi_\textrm{rec}(0,0,x,y)$. (i) $\psi_\textrm{rec}(x,y,x,y)$ without Fourier transform of the original wave function. Panels (d)--(i) are normalized individually.
		}
		\label{figcase1}
	\end{figure}
	
	\subsection{The first case}\label{sec3A}
	The first case is that the positions of the two photons have a good spatial correlation. The wave function has a Gaussian--Schell model-like form,
	\begin{equation}\label{corstate}
		\psi(\mathbf{r}_1,\mathbf{r}_2)=\exp\left[-a(|\mathbf{r}_1|^2+|\mathbf{r}_2|^2)-b|\mathbf{r}_1-\mathbf{r}_2|^2\right],
	\end{equation}
	where $a$ characterizes the size of the light spot and $b$ determines the correlation strength. Photon pairs produced by SPDC roughly (but not exactly \cite{corrreview}) have this form of spatial state. When $a=0$ and $b\rightarrow\infty$, it becomes exactly the EPR state \cite{EPR}. Here $a=1~\textrm{mm}^{-2}$ and $b=1000~\textrm{mm}^{-2}$. path 1 is Fourier transformed and path 2 is added by a phase pattern $\phi_\textrm{add}$ as shown in Fig.~\ref{figcase1} (a) such that
	\begin{equation}\label{corfft}
		\tilde{\psi}(\mathbf{K}_1,\mathbf{r}_2)=e^{-\frac{|\mathbf{K}_1|^2+4ib\mathbf{K}_1\cdot\mathbf{r}_2+4a(a+2b)|\mathbf{r}_2|^2}{4(a+b)}+i\phi_\textrm{add}(\mathbf{r}_2)},
	\end{equation}
	where $\mathbf{K}_1=2\pi\mathbf{r}_1/(\lambda f)$ and the spatial inversion of path 2 from the $4f$ system is omitted. We use $10^9$ photon pairs in each measurement, and slices of the reconstructed and inverse Fourier transformed wave function $\psi_\textrm{rec}(\mathbf{r}_1,\mathbf{r}_2)=\psi_\textrm{rec}(x_1,y_1,x_2,y_2)$ are shown as $\psi_\textrm{rec}(x,y,x,y)$, $\psi_\textrm{rec}(x,0,y,0)$, $\psi_\textrm{rec}(0,x,0,y)$, $\psi_\textrm{rec}(x,y,0,0)$ and $\psi_\textrm{rec}(0,0,x,y)$ in Figs.~\ref{figcase1} (d)--(h). We also reconstructed the single-photon wave function using the marginal intensity distributions of path 1 and 2, which are shown in Figs.~\ref{figcase1} (b) and (c). The result of path 2 is faint compared with the CWS result $\psi_\textrm{rec}(x,y,x,y)$. We can observe the correlation of $\mathbf{r}_1$ and $\mathbf{r}_2$ from Fig.~\ref{figcase1} (e) and (f), and point-like conditional intensity distribution from Figs.~\ref{figcase1} (g) and (h). After inverse FT, the finite size of the ROI broadens the peak width in Figs.~\ref{figcase1} (e)--(h), and may cause the amplitude distribution to deviate from the Gaussian shape.
	
	\begin{figure}[b]
		\centering
		\includegraphics[width=.48\textwidth]{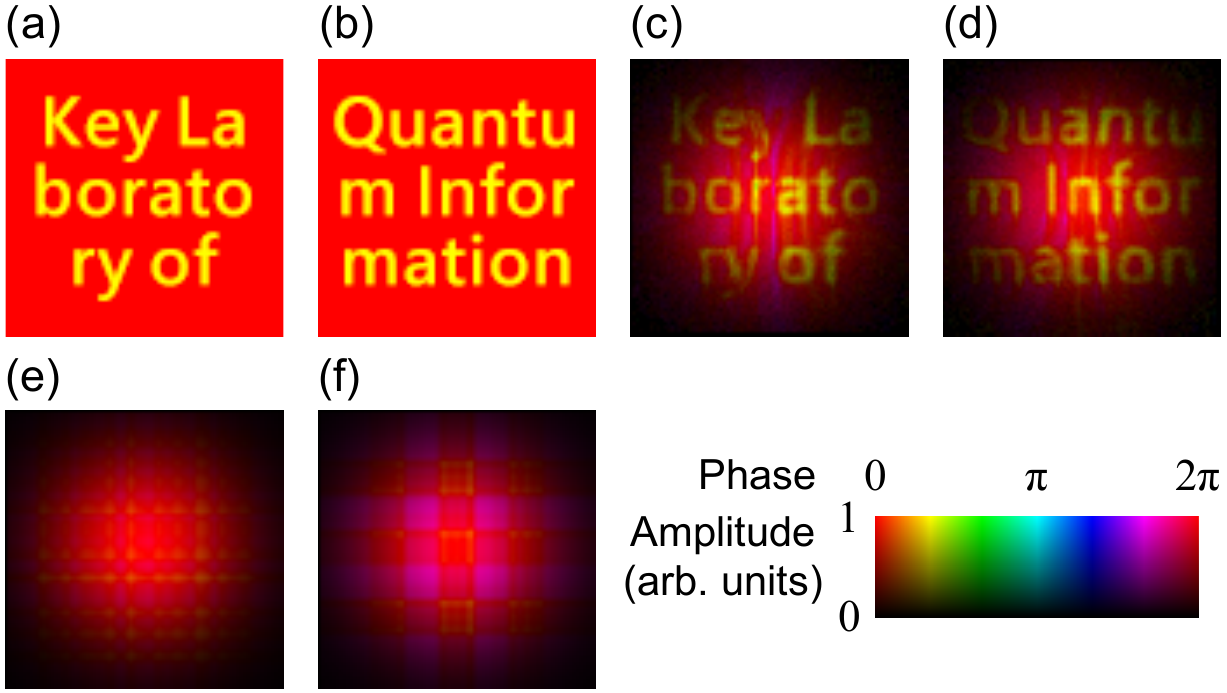}
		\caption{The wave function reconstruction result of the second case of CWS numerical simulation (see text), where two phase patterns are added on $(x_1,x_2)$ and $(y_1,y_2)$ respectively of an uncorrelated Gaussian state. (a) and (b) The phase pattern $\phi_x(x_1,x_2)$ and $\phi_y(y_1,y_2)$. (c) and (d) Slices of the reconstructed wave function $\psi_\textrm{rec}(x,0,y,0)$ and $\psi_\textrm{rec}(0,x,0,y)$. The hue values are added by a constant to improve the readability which does not affect the wave function. (e) and (f) The marginal reconstruction of paths 1 and 2.
		}
		\label{figcase2}
	\end{figure}
	
	\subsection{The second case}
	The phase pattern added on one path is still faintly visible by single path detection. However, for a general correlated phase distribution, this attempt will fail. In the second case, two phase patterns are added on the $x$ and $y$ coordinates respectively. The patterns $\phi_x(x_1,x_2)$ and $\phi_y(y_1,y_2)$ are shown in Figs.~\ref{figcase2} (a) and (b) respectively. The wave function is 
	\begin{align}\label{c2state}
		\psi(\mathbf{r}_1,\mathbf{r}_2)=\exp\big\{&-a(|\mathbf{r}_1|^2+|\mathbf{r}_2|^2)\nonumber\\
		&+i[\phi_x(x_1,x_2)+\phi_y(y_1,y_2)]\big\},
	\end{align}
	where $a=2~\textrm{mm}^{-2}$. This time we do not perform FT (two $4f$ systems on both paths) and the ROI width is $2~\textrm{mm}$. After data generation and processing, reconstructed phase profiles $\psi_\textrm{rec}(x,0,y,0)$ and $\psi_\textrm{rec}(0,x,0,y)$ are shown in Figs.~\ref{figcase2} (c) and (d), where the patterns are revealed. The marginal reconstruction of paths 1 and 2 providing little information are shown in Figs.~\ref{figcase2} (e) and (f). However, this state is difficult to prepare experimentally, so the capability of CWS to detect such a state can only be demonstrated by numerical simulation at present.
	
	\section{Discussions}
	
	Nonclassical lights have advantages in both fundamental \cite{corrreview} and applied physics. With our proposed method, the phase information hidden in the correlation can be measured in a way of adaptive optics, leading to many applications in continuous-variable quantum information processing \cite{cvq}, including the aberration cancellation in free-space quantum communication and experimental studies of biphoton propagation \cite{wolf,Chan2007}. For quantum simulation \cite{simuRMP}, the use of multiphoton spatial mode facilitates the simulation of more complex quantum systems, and this method can be employed to measure the final state. Note that no multiphoton interference is involved here, so the photons can be spatially separated and detected with classical communication. Besides WMWS, the SHWS lens array can be used, which provides a higher efficiency, but the spatial resolution will be reduced \cite{Zheng2021,Zheng2022}. We also expect some direct measurement schemes \cite{Lundeen11,Shi2015} of the multiphoton spatial wave function to be developed some day.
	
	The momentum weak value in Eq.~\eqref{weakvalueeqn} is also related to the Bohmian velocity $\mathbf{v}_j=\operatorname{Re}\langle\hat{\mathbf{p}}_j\rangle_\textrm{w}/m$ \cite{Bohm1,Kocsis2011}, where the mass of photon $m=h/(c\lambda)$ ($h=2\pi\hbar$ and $c$ is the speed of light). In 2013, Braverman and Simon proposed the idea of using a WMWS prototype at one path to study the nonlocality of Bohmian mechanics \cite{Braverman2013}. In relevant experiments, one of the photons is projected to a polarization state \cite{Mahler2016,Xiao2017}. With the CWS in our work, the Bohmian velocity of two photons can be measured with the same setup, and more phenomena like the bidirectional influence between the photons can be experimentally explored in the future.
	
	We should note that wavefront sensing is not a universal method to measure the complex-valued wave function. For example, when the wave function has zeros, the phase step cannot be properly detected. For Laguerre--Gaussian beams with a phase singularity, special algorithms are necessary \cite{Fried2001}. Similarly, these applies to two-photon wave functions. Nevertheless, this method is, in principle, more general than some existing ones \cite{diffract,refint}.
	
	As for choosing one Fourier lens or a $4f$ system, when the wave function has a high position correlation (the probability distribution is almost restricted in a curve of the 4D space whose thickness is comparable to or less than the displacement length of the SP), the weak measurement condition is destroyed. A reconstruction without FT ($10^7$ photon pairs for each measurement) in the first case is shown in Fig.~\ref{figcase1} (i), where the pattern is only faintly visible. FT is needed to project photons on one path into the momentum space to ensure the validity of weak measurement, and thus the phase can be properly reconstructed. To further quantify whether to perform FT, we consider the first case in the numerical simulation with an arbitrary $b$, and calculate the conditional probability distribution of photon 1 when photon 2 is at $\mathbf{r}_2=0$ from Eq.~\eqref{corstate} or Eq.~\eqref{corfft},
	\begin{gather}\label{cidb}
		|\psi(\mathbf{r}_1|\mathbf{r}_2=0)|^2=\exp\left[-2(a+b)|\mathbf{r}_1|^2\right],\nonumber\\
		\big|\tilde{\psi}(\mathbf{r}_1|\mathbf{r}_2=0)\big|^2=\exp\left[-\frac{2\pi^2|\mathbf{r}_1|^2}{\lambda^2f^2(a+b)}\right].
	\end{gather}
	When $\lambda$ and $f$ take the values in Section \ref{sec3} and $a=1~\textrm{mm}^{-2}$, the widths of the two conditional probability distributions are equal when $b\approx19.6~\textrm{mm}^{-2}$, and the corresponding full width at half maximum (FWHM) values are approximately $266~\textrm{\textmu m}$. The weak measurement condition (the displacement induced by the SP is $50$~\textmu m) is not severely violated. When $b$ is greater than this value, FT is recommended; otherwise we should use the $4f$ system.
	
	As for experiments, the traditional time-consuming method is by raster scanning the point detectors \cite{ghostimag,Howell2004}. Defienne \emph{et al.}\ made pioneering works on the joint probability distribution acquisition, by taking multiple frames using an electron-multiplying charge-coupled device (EMCCD) \cite{Defienne2018,Reichert2018,Reichert2018PRA,Defienne20182,polarent}. However, its signal-to-noise ratio hinders its practicality when the conditional probability distribution is too wide \cite{OAMEMCCD}, corresponding to the two numerical simulation cases. Recently, it was shown that a single-photon-sensitive time-stamping camera named Tpx3Cam is able to perform spatially resolved coincidence photon counting \cite{arxiv2023}. Our work shows the necessity to develop more efficient photon-counting cameras for future quantum optics and quantum communication technologies.
	
	\section*{Acknowledgments}
	
	This work was supported by the Innovation Program for Quantum Science and Technology (Grant No.\ 2021ZD0301400), the National Natural Science Foundation of China (Grants No.\ 61725504, No.\ 11821404, and No.\ U19A2075), Anhui Initiative in Quantum Information Technologies (Grants No.\ AHY020100 and No.\ AHY060300), and the Fundamental Research Funds for the Central Universities (Grants No.\ WK2030380017 and No.\ WK5290000003).
	
	\appendix
	\begin{widetext}
		\section{Details of first-order approximations}\label{appA}
		
		Without approximation, in the $k_x$ measurement, the conditional intensity distributions are
		\begin{align}\label{condint}
			I_{1L}(\mathbf{r}_1|\mathbf{r}_2)&\propto A^2_{++}+A^2_{+-}+A^2_{-+}+A^2_{--}-2\left[A_{++}A_{-+}\sin(\phi_{-+}-\phi_{++})+A_{+-}A_{--}\sin(\phi_{--}-\phi_{+-})\right],\nonumber\\
			I_{1R}(\mathbf{r}_1|\mathbf{r}_2)&\propto A^2_{++}+A^2_{+-}+A^2_{-+}+A^2_{--}+2\left[A_{++}A_{-+}\sin(\phi_{-+}-\phi_{++})+A_{+-}A_{--}\sin(\phi_{--}-\phi_{+-})\right].
		\end{align}
		Assuming the functions in a small region centered by point $(\mathbf{r}_1,\mathbf{r}_2)$ can be approximated by, for example,
		\begin{equation}
			A(\mathbf{r}_1+\Delta\mathbf{r}_1,\mathbf{r}_2+\Delta\mathbf{r}_2)\approx A(\mathbf{r}_1,\mathbf{r}_2)+\nabla_1 A(\mathbf{r}_1,\mathbf{r}_2)\cdot\Delta\mathbf{r}_1+\nabla_2 A(\mathbf{r}_1,\mathbf{r}_2)\cdot\Delta\mathbf{r}_2
		\end{equation}
		and $l$ characterizing the displacement length of the Savart plate is a small quantity, we have $\phi_{-\pm}-\phi_{+\pm}\approx2lk_{1x}(\mathbf{r}_1-l\mathbf{e}_y,\mathbf{r}_2-l\mathbf{e}_y)$ and
		\begin{gather}
			A^2_{++}+A^2_{+-}+A^2_{-+}+A^2_{--}\approx 4\left(A^2-2A\frac{\partial A}{\partial y_1}l-2A\frac{\partial A}{\partial y_2}l\right)\approx 4A(\mathbf{r}_1-l\mathbf{e}_y,\mathbf{r}_2-l\mathbf{e}_y)^2,\nonumber\\
			A_{++}A_{-+}+A_{+-}A_{--}\approx 2A(\mathbf{r}_1-l\mathbf{e}_y,\mathbf{r}_2-l\mathbf{e}_y)^2,
		\end{gather}
		where second-order small quantities have been omitted. So,
		\begin{align}\label{condintres1}
			I_{1L}(\mathbf{r}_1|\mathbf{r}_2)&\propto A(\mathbf{r}_1-l\mathbf{e}_y,\mathbf{r}_2-l\mathbf{e}_y)^2\left\{1-\sin[2lk_{1x}(\mathbf{r}_1-l\mathbf{e}_y,\mathbf{r}_2-l\mathbf{e}_y)]\right\},\nonumber\\
			I_{1R}(\mathbf{r}_1|\mathbf{r}_2)&\propto A(\mathbf{r}_1-l\mathbf{e}_y,\mathbf{r}_2-l\mathbf{e}_y)^2\left\{1+\sin[2lk_{1x}(\mathbf{r}_1-l\mathbf{e}_y,\mathbf{r}_2-l\mathbf{e}_y)]\right\}.
		\end{align}
		
		\section{A derivation with weak measurement}\label{appB}
		
		The derivation method does not necessarily involve the concept of weak value. Now we rearrange it under the framework of weak measurement and weak value, since we named this wavefront sensor with ``weak measurement.'' The action of two Savart plates (not rotated in $k_x$ measurement) is described by a unitary operator $\hat{U}_\textrm{SP}$ acting on the initial state
		\begin{align}
			\hat{U}_\textrm{SP}|\tilde{\psi}\rangle|H_1\rangle|H_2\rangle&=e^{-i\hat{\mathbf{p}}_2\cdot\mathbf{l}_-|A_2\rangle\langle A_2|/\hbar}
			e^{-i\hat{\mathbf{p}}_2\cdot\mathbf{l}_+|D_2\rangle\langle D_2|/\hbar}
			e^{-i\hat{\mathbf{p}}_1\cdot\mathbf{l}_-|A_1\rangle\langle A_1|/\hbar}
			e^{-i\hat{\mathbf{p}}_1\cdot\mathbf{l}_+|D_1\rangle\langle D_1|/\hbar}
			|\tilde{\psi}\rangle|H_1\rangle|H_2\rangle\nonumber\\
			&=\exp\left[-\frac{i}{\hbar}\hat{\mathbf{p}}_1\cdot\left(\mathbf{l}_+|D_1\rangle\langle D_1|+\mathbf{l}_-|A_1\rangle\langle A_1|\right)-\frac{i}{\hbar}\hat{\mathbf{p}}_2\cdot\left(\mathbf{l}_+|D_2\rangle\langle D_2|+\mathbf{l}_-|A_2\rangle\langle A_2|\right)\right]|\tilde{\psi}\rangle|H_1\rangle|H_2\rangle\nonumber\\
			&=\exp\left[-\frac{il}{\hbar}\left(\hat{p}_{1y}|H_1\rangle\langle H_1|+\hat{p}_{1x}|V_1\rangle\langle H_1|+\hat{p}_{2y}|H_2\rangle\langle H_2|+\hat{p}_{2x}|V_2\rangle\langle H_2|\right)\right]|\tilde{\psi}\rangle|H_1\rangle|H_2\rangle\nonumber\\
			&\approx|\tilde{\psi}\rangle|H_1\rangle|H_2\rangle-\frac{il}{\hbar}\left(\hat{p}_{1y}|\tilde{\psi}\rangle|H_1\rangle|H_2\rangle+\hat{p}_{1x}|\tilde{\psi}\rangle|V_1\rangle|H_2\rangle+\hat{p}_{2y}|\tilde{\psi}\rangle|H_1\rangle|H_2\rangle+\hat{p}_{2x}|\tilde{\psi}\rangle|H_1\rangle|V_2\rangle\right),
		\end{align}
		where first-order approximations have been used from the start, and the terms in the exponent corresponding to the actions of four individual birefringent crystals commute with each other. Then the state is postselected into given circular polarizations and positions $|\mathbf{r}_1,\mathbf{r}_2\rangle|{^L_R}_1\rangle|{^L_R}_2\rangle$, that is,
		\begin{align}
			\langle\mathbf{r}_1,\mathbf{r}_2|\langle{^L_R}_1|\langle{^L_R}_2|\hat{U}_\textrm{SP}|\tilde{\psi}\rangle|H_1\rangle|H_2\rangle&\approx\frac{1}{2}\left[\langle\mathbf{r}_1,\mathbf{r}_2|\tilde{\psi}\rangle+\frac{l}{\hbar}\langle\mathbf{r}_1,\mathbf{r}_2|(\mp_1 \hat{p}_{1x}\mp_2 \hat{p}_{2x})|\tilde{\psi}\rangle-i\frac{l}{\hbar}\langle\mathbf{r}_1,\mathbf{r}_2|(\hat{p}_{1y}+\hat{p}_{2y})|\tilde{\psi}\rangle\right]\nonumber\\
			&=\frac{1}{2}\langle\mathbf{r}_1,\mathbf{r}_2|\tilde{\psi}\rangle\left[1+\frac{l}{\hbar}\left(\mp_1\langle\hat{p}_{1x}\rangle_\textrm{w}\mp_2\langle\hat{p}_{2x}\rangle_\textrm{w}-i\langle \hat{p}_{1y}+\hat{p}_{2y}\rangle_\textrm{w}\right)\right],
		\end{align}
		where the weak values $\langle\hat{A}\rangle_\textrm{w}=\langle\boldsymbol{r}_1,\boldsymbol{r}_2|\hat{A}|\tilde{\psi}\rangle/\langle\boldsymbol{r}_1,\boldsymbol{r}_2|\tilde{\psi}\rangle$. The real part of the momentum weak value $\operatorname{Re}\langle\hat{p}_{jx}\rangle_\textrm{w}$ equals $\hbar k_{jx}$. The four joint intensity distributions $I_{LL},I_{LR},I_{RL},I_{RR}$ are proportional to the probability of the projection
		\begin{align}
			\left|\langle\mathbf{r}_1,\mathbf{r}_2|\langle{^L_R}_1|\langle{^L_R}_2|\hat{U}_\textrm{SP}|\tilde{\psi}\rangle|H_1\rangle|H_2\rangle\right|^2&\approx\frac{1}{4}\left|\langle\mathbf{r}_1,\mathbf{r}_2|\tilde{\psi}\rangle\right|^2\left|1+\frac{l}{\hbar}\left(\mp_1\langle\hat{p}_{1x}\rangle_\textrm{w}\mp_2\langle\hat{p}_{2x}\rangle_\textrm{w}-i\langle\hat{p}_{1y}+\hat{p}_{2y}\rangle_\textrm{w}\right)\right|^2\nonumber\\
			&\approx\frac{1}{4}\left|\langle\mathbf{r}_1,\mathbf{r}_2|\tilde{\psi}\rangle\right|^2\left[1+\frac{2l}{\hbar}\left(\mp_1\operatorname{Re}\langle\hat{p}_{1x}\rangle_\textrm{w}\mp_2\operatorname{Re}\langle\hat{p}_{2x}\rangle_\textrm{w}\right)\right].
		\end{align}

	So,
	\begin{align}\label{ap1I}
		I_{1{^L_R}}\propto\frac{1}{2}\left|\langle\mathbf{r}_1,\mathbf{r}_2|\tilde{\psi}\rangle\right|^2\left(1\mp\frac{2l}{\hbar}\operatorname{Re}\langle\hat{p}_{1x}\rangle_\textrm{w}\right).
	\end{align}
	Now Eq.~\eqref{ap1I} takes a similar form as Eq.~\eqref{condintres1} except for the absence of the sine operation and the small position deviation, which are reasonable from the first-order approximations.
	
	\section{Generalization to $n$ photons}\label{appC}
	
	In principle, the CWS scheme is extensible to the detection of the joint spatial wave function of $n$ photons, provided the coincidence counting is experimentally realizable, and certain Fourier transforms are needed when the position distribution is highly correlated to ensure the weak measurement. Here we still use the clearer weak measurement approach. The initial state is $|\tilde{\psi}\rangle|H_1\rangle|H_2\rangle\cdots|H_n\rangle=|\tilde{\psi}\rangle\bigotimes_{j=1}^n|H_j\rangle$, and the action of $n$ Savart plates is described by (terms ending with $\langle V_j|$ are omitted)
	\begin{equation}
		\hat{U}_\textrm{SP}=\exp\left[-\frac{il}{\hbar}\sum_{j=0}^n\left(\hat{p}_{jy}|H_j\rangle+\hat{p}_{jx}|V_j\rangle\right)\langle H_j|\right].
	\end{equation}
	Acted by $\hat{U}_\textrm{SP}$ and postselected by the positions and polarizations $|\mathbf{r}_1,\mathbf{r}_2,\ldots,\mathbf{r}_n\rangle\bigotimes_{j=1}^n|{^L_R}_j\rangle$, under first-order approximations, the projection probability is
		\begin{equation}
			\left|\langle\mathbf{r}_1,\mathbf{r}_2,\ldots,\mathbf{r}_n|\bigotimes_{j=1}^n\langle{^L_R}_j|\hat{U}_\textrm{SP}|\tilde{\psi}\rangle\bigotimes_{k=1}^n|H_k\rangle\right|^2\approx\frac{1}{2^n}\left|\langle\mathbf{r}_1,\mathbf{r}_2,\ldots,\mathbf{r}_n|\tilde{\psi}\rangle\right|^2\left[1+\sum_{j=0}^n\mp_j\sin\left(\frac{2l}{\hbar}\operatorname{Re}\langle\hat{p}_{jx}\rangle_\textrm{w}\right)\right],
		\end{equation}
		which we denote by $I_{{^L_R}_1,{^L_R}_2,\ldots,{^L_R}_n}(\mathbf{r}_1,\mathbf{r}_2,\ldots,\mathbf{r}_n)$. Note that the sine operation has been deliberately added. Letting $I_{j{^L_R}}(\mathbf{r}_j|\mathbf{r}_\textrm{others})$ be the sum of all the intensities with the $j$th photon postselected to $L$ or $R$ polarization, we have
		\begin{equation}
			I_{j{^L_R}}(\mathbf{r}_j|\mathbf{r}_\textrm{others})\approx\frac{1}{2}\left|\langle\mathbf{r}_1,\mathbf{r}_2,\ldots,\mathbf{r}_n|\tilde{\psi}\rangle\right|^2\left[1\mp\sin\left(\frac{2l}{\hbar}\operatorname{Re}\langle\hat{p}_{jx}\rangle_\textrm{w}\right)\right],
		\end{equation}
		and then we can obtain $k_{jx}(\mathbf{r}_1,\mathbf{r}_2,\ldots,\mathbf{r}_n)$. $k_{jy}$ is measured after rotating all the Savart plates by $90^\circ$. With the phase gradient $\nabla\arg\tilde{\psi}(\mathbf{r}_1,\mathbf{r}_2,\ldots,\mathbf{r}_n)$, the wave function can be reconstructed.
		
		\section{Detection of an individual path}\label{appD}
		
		With coincidence measurement, we can obtain the conditional intensity distribution of photon 1 given the postselected position of photon 2, $I_{1L}(\mathbf{r}_1|\mathbf{r}_2)$ and $I_{1R}(\mathbf{r}_1|\mathbf{r}_2)$. If we do not use the information of photon 2, using Eq.~\eqref{ap1I}, the marginal intensity distribution of photon 1 is the integral over $\mathbf{r}_2$,
		\begin{align}
			I_{1{^L_R}}(\mathbf{r}_1)&=\int d\mathbf{r}_2 I_{1{^L_R}}(\mathbf{r}_1|\mathbf{r}_2)\propto\int d\mathbf{r}_2 \left|\langle\mathbf{r}_1,\mathbf{r}_2|\tilde{\psi}\rangle\right|^2\left(1\mp\frac{2l}{\hbar}\operatorname{Re}\frac{\langle\mathbf{r}_1,\mathbf{r}_2|\hat{p}_{1x}|\tilde{\psi}\rangle}{\langle\mathbf{r}_1,\mathbf{r}_2|\tilde{\psi}\rangle}\right)\nonumber\\
			&=\int d\mathbf{r}_2 \left(\langle\mathbf{r}_1,\mathbf{r}_2|\tilde{\psi}\rangle\langle\tilde{\psi}|\mathbf{r}_1,\mathbf{r}_2\rangle\mp\frac{2l}{\hbar}\operatorname{Re}\langle\mathbf{r}_1,\mathbf{r}_2|\hat{p}_{1x}|\tilde{\psi}\rangle\langle\tilde{\psi}|\mathbf{r}_1,\mathbf{r}_2\rangle\right)=\langle\mathbf{r}_1|\hat{\tilde{\rho}}_1|\mathbf{r}_1\rangle\left(1\mp\frac{2l}{\hbar}\operatorname{Re}\frac{\langle\mathbf{r}_1|\hat{p}_{1x}\hat{\tilde{\rho}}_1|\mathbf{r}_1\rangle}{\langle\mathbf{r}_1|\hat{\tilde{\rho}}_1|\mathbf{r}_1\rangle}\right),
		\end{align}
		where $\hat{\tilde{\rho}}_1=\operatorname{Tr}_2|\tilde{\psi}\rangle\langle\tilde{\psi}|=\int d\mathbf{r}_2\langle\mathbf{r}_2|\tilde{\psi}\rangle\langle\tilde{\psi}|\mathbf{r}_2\rangle$ is the reduced density operator, and the fraction in the bracket is the weak value with a mixed initial state \cite{wvmix}. Similarly, with $\hat{\rho}_2=\operatorname{Tr}_1|\tilde{\psi}\rangle\langle\tilde{\psi}|$ (the tilde symbol is omitted because the Fourier transform does not act on photon 2), we have
		\begin{equation}
			I_{2{^L_R}}(\mathbf{r}_2)=\langle\mathbf{r}_2|\hat{\rho}_2|\mathbf{r}_2\rangle\left(1\mp\frac{2l}{\hbar}\operatorname{Re}\frac{\langle\mathbf{r}_2|\hat{p}_{2x}\hat{\rho}_2|\mathbf{r}_2\rangle}{\langle\mathbf{r}_2|\hat{\rho}_2|\mathbf{r}_2\rangle}\right).
		\end{equation}
		
		If a pure-phase object $\phi_\textrm{add}(\mathbf{r}_2)$ is added in path 2 and the state becomes $|\tilde{\psi}'\rangle=\exp[i\phi_\textrm{add}(\hat{\mathbf{r}}_2)]|\tilde{\psi}\rangle$, we can easily verify $\hat{\tilde{\rho}}'_1=\hat{\tilde{\rho}}_1$. For path 2, $\hat{\rho}'_2=\exp[i\phi_\textrm{add}(\hat{\mathbf{r}}_2)]\hat{\rho}_2\exp[-i\phi_\textrm{add}(\hat{\mathbf{r}}_2)]$, $\langle\mathbf{r}_2|\hat{\rho}'_2|\mathbf{r}_2\rangle=\langle\mathbf{r}_2|\hat{\rho}_2|\mathbf{r}_2\rangle$ is real, and
		\begin{align}
			\langle\mathbf{r}_2|\hat{p}_{2x}\hat{\rho}'_2|\mathbf{r}_2\rangle
			&=e^{-i\phi_\textrm{add}(\mathbf{r}_2)}\langle\mathbf{r}_2|\hat{p}_{2x}e^{i\phi_\textrm{add}(\hat{\mathbf{r}}_2)}\hat{\rho}_2|\mathbf{r}_2\rangle
			=e^{-i\phi_\textrm{add}(\mathbf{r}_2)}\left\{-i\hbar\frac{\partial}{\partial x'_2}\left[e^{i\phi_\textrm{add}(\mathbf{r}'_2)}\langle\mathbf{r}'_2|\hat{\rho}_2|\mathbf{r}_2\rangle\right]\middle\}\right|_{\mathbf{r}'_2=\mathbf{r}_2}\nonumber\\
			&=\left(-i\hbar\frac{\partial}{\partial x'_2}\langle\mathbf{r}'_2|\hat{\rho}_2|\mathbf{r}_2\rangle\middle)\right|_{\mathbf{r}'_2=\mathbf{r}_2}+\hbar\langle\mathbf{r}_2|\hat{\rho}_2|\mathbf{r}_2\rangle\frac{\partial\phi_\textrm{add}(\mathbf{r}_2)}{\partial x_2}
			=\langle\mathbf{r}_2|\hat{p}_{2x}\hat{\rho}_2|\mathbf{r}_2\rangle+\hbar\langle\mathbf{r}_2|\hat{\rho}_2|\mathbf{r}_2\rangle\frac{\partial\phi_\textrm{add}(\mathbf{r}_2)}{\partial x_2}.
		\end{align}
		So,
		\begin{equation}
			\operatorname{Re}\frac{\langle\mathbf{r}_2|\hat{p}_{2x}\hat{\rho}'_2|\mathbf{r}_2\rangle}{\langle\mathbf{r}_2|\hat{\rho}'_2|\mathbf{r}_2\rangle}=\operatorname{Re}\frac{\langle\mathbf{r}_2|\hat{p}_{2x}\hat{\rho}_2|\mathbf{r}_2\rangle}{\langle\mathbf{r}_2|\hat{\rho}_2|\mathbf{r}_2\rangle}+\hbar\frac{\partial\phi_\textrm{add}(\mathbf{r}_2)}{\partial x_2},
		\end{equation}
		which means the difference in the value before and after phase addition can reconstruct the added phase distribution \cite{Zheng2021}. However, this only applies when $l$ is sufficiently small to ensure the weak measurement. In most cases, good correlation causes the reduced density operator to have a sharp peak in the position space, i.e., $\langle\mathbf{r}_2|\hat{\rho}_2|\mathbf{r}'_2\rangle\approx0$ when $|\mathbf{r}_2-\mathbf{r}'_2|\sim l$, and the measured $\mathbf{k}_2$ is almost zero no matter what phase is added. We use the strict approach to illustrate this point for a single photon at the initial state $\hat{\rho}|H\rangle\langle H|$. The action of the Savart plate is $\hat{U}_\textrm{SP}=\exp(-i\hat{\mathbf{p}}\cdot\mathbf{l}_-|A\rangle\langle A|/\hbar)
		\exp(-i\hat{\mathbf{p}}\cdot\mathbf{l}_+|D\rangle\langle D|/\hbar)$, and we have
		\begin{equation}
			\langle\mathbf{r}|\langle{^L_R}|\hat{U}_\textrm{SP}\hat{\rho}|H\rangle\langle H|\hat{U}_\textrm{SP}^\dagger|\mathbf{r}\rangle|{^L_R}\rangle=
			\frac{1}{4}\left(\langle\mathbf{r}-\mathbf{l}_+|\hat{\rho}|\mathbf{r}-\mathbf{l}_+\rangle+\langle\mathbf{r}-\mathbf{l}_-|\hat{\rho}|\mathbf{r}-\mathbf{l}_-\rangle\mp i\langle\mathbf{r}-\mathbf{l}_+|\hat{\rho}|\mathbf{r}-\mathbf{l}_-\rangle\pm i\langle\mathbf{r}-\mathbf{l}_-|\hat{\rho}|\mathbf{r}-\mathbf{l}_+\rangle\right).
		\end{equation}
		If the last two terms are zero, the postselection probability is the same for $L$ and $R$ polarization, and $k_{x}=0$.
		
		Now we use the wave function in Eq.~\eqref{corstate} as a typical two-photon correlated wave function to calculate the reduced density operator. After calculation,
		\begin{equation}
			\langle\mathbf{r}|\hat{\rho}_{ab2}|\mathbf{r}'\rangle\propto\int d\mathbf{r}_1 \psi_{ab}(\mathbf{r}_1,\mathbf{r})\psi^*_{ab}(\mathbf{r}_1,\mathbf{r}')\propto\exp\left[-\frac{2a(a+2b)\left(|\mathbf{r}'|^2+|\mathbf{r}|^2\right)+b^2|\mathbf{r}'-\mathbf{r}|^2}{2(a+b)}\right].
		\end{equation}
		When $a\ll b$, the weak value can be correctly measured if $l\ll 1/\sqrt{b}$, and the added phase will be almost undetectable if $l>\sqrt{2/b}$.
		
		\section{Phase reconstruction algorithm}\label{appE}
		
		The goal of the reconstruction algorithm is to realize the line integral in Eq.~\eqref{lineint}, when the acquired $\boldsymbol{k}_1,\boldsymbol{k}_2$ distributions contain errors, which will become significant if we directly expand the phase values from an initial point sequentially. The basic idea of the algorithm is as follows:
		
		(1) Set the phase of the point with the maximum intensity (named initial point) to be zero, and other points to be ``unfilled.''
		
		(2) A queue is used in the algorithm. The initial point is enqueued.
		
		(3) Multiple computer threads attempt to dequeue a point from the queue. If one thread finds the queue is empty, it goes to step (5). Each thread checks how many unfilled points neighbor the point. These points are filled in probabilistically, that is, only part of them are filled and enqueued. If there exists neighboring points left unfilled, the previously dequeued point is enqueued again. If we denote a dequeued filled point as $(m_1,n_1,m_2,n_2)$ ($m_1,n_1,m_2,n_2$ are integers), the basic expansion formulas are, for example,
		\begin{gather}
			\phi(m_1-1,n_1,m_2,n_2)=\phi(m_1,n_1,m_2,n_2)-k_{1x}(m_1-1,n_1,m_2,n_2)\varDelta,\nonumber\\
			\phi(m_1+1,n_1,m_2,n_2)=\phi(m_1,n_1,m_2,n_2)+k_{1x}(m_1,n_1,m_2,n_2)\varDelta.
		\end{gather}
		The probability of filling is smaller in regions where the intensity is lower, as errors in $k_x,k_y$ are larger in these areas.
		
		(4) The thread goes back to step (3).
		
		(5) When all the threads report the queue is empty, one reconstruction process is done. The phase distribution is recorded. If not, the threads go back to step (3).
		
		(6) To reduce the error, go back to step (1) and repeat the process several times [25 times in the main numerical simulations, and 50 times in Fig.~\ref{figcase1} (i)]. The average phase value is calculated from the recorded phases, and is taken as the final phase distribution.
		
		(7) Finally, the wave function is calculated as $\psi_\textrm{rec}=\sqrt{I_{1L}+I_{1R}}\exp(i\phi)$. The inverse FT may be performed.
		
		The source code of the numerical simulation program \cite{SourceCode} provides all the details of this algorithm. Note that the numerical simulation above took up a computer memory space of a few gigabytes. If the spatial resolution or the number of photons $n$ increases, the memory usage soar up violently.
	\end{widetext}


\begin{thebibliography}{99}
\bibitem{quanoptnote}
		The wave function $\psi(\mathbf{r}_1,\mathbf{r}_2)=\langle\mathbf{r}_1,\mathbf{r}_2|\psi\rangle$ is actually $\langle0|E^{(+)}(\mathbf{r}_1,t)E^{(+)}(\mathbf{r}_2,t)|\varPsi\rangle$ in quantum optics, where $|0\rangle$ is the vacuum state, $E^{(+)}$ is the electric field operator containing the annihilation operators, the $z_1,z_2,t$ values are given, and $|\varPsi\rangle$ is the state of the quantized field with two photons.
\end{thebibliography}
\end{document}